\documentclass{aa}  

\usepackage{graphicx}
\usepackage[varg]{txfonts}
\usepackage{natbib}
\usepackage{hyperref}

\begin{document} 
   \title{Observations of water with {\it Herschel}/HIFI toward the high-mass protostar AFGL 2591\thanks{{\it Herschel} is an ESA space observatory with science instruments provided by European-led Principal Investigator consortia and with important participation from NASA.}}

   \author{Y. Choi\inst{1,2},
           F. F. S. van der Tak\inst{2,1},
           E. F. van Dishoeck\inst{3,4},
           F. Herpin\inst{5,6},
           F. Wyrowski\inst{7}}

   \institute{Kapteyn Astronomical Institute, University of Groningen, P.O. Box 800, 9700 AV, Groningen, The Netherlands \\
              \email{y.choi@astro.rug.nl} \and
              SRON Netherlands Institute for Space Research, P.O. Box 800, 9700 AV, Groningen, The Netherlands \and
              Leiden Observatory, Leiden University, P.O. Box 9513, 2300 RA Leiden, The Netherlands \and
              Max Planck Institut f\"{u}r Extraterrestrische Physik, Giessenbachstrasse 1, 85748 Garching, Germany \and
               Universit{\'e} de Bordeaux, Observatoire Aquitain des Sciences de l'Univers, 2 rue de l'Observatoire, BP 89, F-33270 Floirac Cedex, France \and
              CNRS, LAB, UMR 5804, Laboratoire d'Astrophysique de Bordeaux, 2 rue de l'Observatoire, BP 89, F-33270 Floirac Cedex, France \and
              Max Planck Institut f\"{u}r Radioastronomie, Auf dem H\"{u}gel 69, 53121 Bonn, Germany}
   \date{Received ; accepted }


  \abstract
   {Water is an important chemical species in the process of star formation, and a sensitive tracer of
    physical conditions in star-forming regions because of its rich line spectrum and large abundance
    variations between hot and cold regions.}
    {We use spectrally resolved observations of rotational lines of H$_2$O and its isotopologs to constrain
     the physical conditions of the water emitting region toward the high-mass protostar AFGL 2591.}
    {{\it Herschel}-HIFI spectra from 552 up to 1669 GHz show emission and absorption in 14 lines of
     H$_2$O, H$_2$$^{18}$O, and H$_2$$^{17}$O. We decompose the line profiles into contributions from the protostellar envelope,
     the bipolar outflow, and a foreground cloud. We use analytical estimates and rotation diagrams to estimate excitation
     temperatures and column densities of H$_2$O in these components.
     Furthermore, we use the non-LTE radiative transfer code RADEX
     to estimate the temperature and volume density of the H$_2$O emitting gas.}
    { Assuming LTE, we estimate an excitation temperature of $\sim$42 K and a column density of $\sim$2$\times$ 10$^{14}$ cm$^{-2}$ for the envelope
     and $\sim$45 K and 4$\times$10$^{13}$ cm$^{-2}$ for the outflow, in beams of 4$\arcsec$ and 30$\arcsec$, respectively.
     Non-LTE models indicate a kinetic temperature of $\sim$60$-$230 K and a volume density of 7$\times$10$^6$$-$10$^8$ cm$^{-3}$ for the envelope, and a kinetic temperature of $\sim$70$-$90 K and a gas density of $\sim$10$^7$$-$10$^8$ cm$^{-3}$ for the outflow.
     The ortho/para ratio of the narrow cold foreground absorption is lower than three ($\sim$1.9$\pm$0.4), suggesting a low temperature.
     In contrast, the ortho/para ratio seen in absorption by the outflow is about 3.5$\pm$1.0, as expected for warm gas.}
   {The water abundance in the outer envelope of AFGL 2591 is $\sim$10$^{-9}$ for a source size of 4$\arcsec$, similar to the low values found for other high-mass and low-mass protostars, 
suggesting that this abundance is constant during the embedded phase of high-mass star formation.
    The water abundance in the outflow is $\sim$10$^{-10}$ for a source size of 30$\arcsec$, which is $\sim$10$\times$ lower than in the envelope and in the outflows of high-mass and low-mass protostars. 
Since beam size effects can only increase this estimate by a factor of 2, we suggest that the water in the AFGL 2591 outflow is affected by dissociating UV radiation due to the low extinction in the outflow lobe.
}


 \authorrunning{Y. Choi et al.}
 \titlerunning{Water in high-mass protostar AFGL 2591 with {\it Herschel}/HIFI}

   \maketitle
%

\section{Introduction}

Massive stars play a major role in the interstellar energy budget and the shaping of the galactic environment.
However, the formation of high-mass stars is not well understood for several reasons: they are rare, they have a short evolution time scale, they are born deeply embedded,
and they are far from us.
The water molecule is thought to be a sensitive tracer of physical conditions and
dynamics in star-forming regions because of its large abundance variations between hot and cold regions.
Water is also an important reservoir of oxygen and therefore a crucial ingredient in the chemistry of oxygen-bearing molecules.
In the surroundings of embedded protostars, water can be formed by three different mechanisms \citep[see][for a review]{vDishoeck13}.
First, in molecular clouds, water may be formed in the gas phase by ion-molecule chemistry
through dissociative recombination of H$_3$O$^+$. Second, in cold and dense cores, on the surfaces of cold dust grains, O and H atoms may combine to form water-rich ice mantles.
These mantles will evaporate when the grains are heated to $\sim$100 K by protostellar radiation or sputtered by outflow shocks.
Third, in gas with temperatures above 300 K, reactions of O and OH with H$_2$ drive all gas-phase oxygen into water.
Such high temperatures may occur very close to the stars, or near outflow shocks.
Therefore, measurement of the water abundance is a step towards understanding the star formation process.

\begin{table*}
\caption{Observed lines.}
\label{table:1}
\centering
\begin{tabular}{l l c c c c c c c c c c}
\hline\hline
Molecule & Transition  & Obsid & $\nu$ & $E_{\rm up}$ & $T_{\rm sys}$ & $t_{\rm int}$ & Beam     & $\eta_{\rm mb}$  & $\delta v $  & rms \\
         &             &         & (GHz) & (K)          & (K)           & (min)         & (\arcsec) &                  & (MHz) & (mK) \\
\hline
o-H$_2$$^{18}$O\tablefootmark{a}  & $1_{10}-1_{01}$ & 1342210763 & 547.676 & 60.5 & 74 & 33 & 38.7 & 0.75 & 0.48 & 14 \\
o-H$_2$$^{17}$O                                & $1_{10}-1_{01}$ & 1342192360 & 552.020 & 61.0  &  72  & 2.2 & 38.0 & 0.75 & 0.48  & 33 \\
o-H$_2$O\tablefootmark{a}                & $1_{10}-1_{01}$ & 1342210763 & 556.936 & 61.0 & 74 & 33 & 38.0 & 0.75 & 0.24 & 14  \\
p-H$_2$O                                               & $2_{11}-2_{02}$ & 1342192335 & 752.033 & 136.9 & 178  & 3.9 & 28.2 & 0.75 & 0.24 & 83 \\
p-H$_2$O                                               & $2_{02}-1_{11}$ & 1342195019 & 987.926 & 100.8 & 371  & 6.3 & 21.3 & 0.74 & 0.24 & 115 \\
p-H$_2$$^{18}$O                                 & $2_{02}-1_{11}$ & 1342195020 & 994.675 & 100.8 & 276  & 7.8 & 21.3 & 0.74 & 0.48 & 66 \\
o-H$_2$$^{18}$O                                 & $3_{12}-3_{03}$ & 1342194796 & 1095.627 & 249.4 & 373  & 27  & 19.2 & 0.74 & 0.48 & 51 \\
o-H$_2$O                                               & $3_{12}-3_{03}$ & 1342194796 & 1097.365 & 249.4 & 373  & 27  & 19.2 & 0.74 & 0.24 & 51 \\
p-H$_2$$^{18}$O                                 & $1_{11}-0_{00}$ & 1342194795, 1342197973 & 1101.698 & 53.4  & 350  & 56  & 19.2 & 0.74 & 0.48 & 33\\
p-H$_2$$^{17}$O                                 & $1_{11}-0_{00}$ & 1342194796 & 1107.166 & 53.4  & 373  & 27  & 19.1 & 0.74 & 0.48 & 42 \\
p-H$_2$O                                               & $1_{11}-0_{00}$ & 1342194795, 1342197973 & 1113.343 & 53.4  & 390  & 57  & 19.0 & 0.74 & 1.1 & 33 \\
o-H$_2$O                                               & $2_{21}-2_{12}$ & 1342192579 & 1661.007 & 194.1 & 1416 & 16  & 12.8 & 0.71 & 1.1 & 166 \\
o-H$_2$$^{17}$O                                 & $2_{12}-1_{01}$ & 1342192579 & 1662.464 & 113.6 & 1417 & 17  & 12.8 & 0.71 & 1.1 & 103 \\
o-H$_2$O                                               & $2_{12}-1_{01}$ & 1342192579 & 1669.904 & 114.4 & 1417 & 17  & 12.7 & 0.71 & 1.1 & 95 \\
\hline
\end{tabular}
\tablefoot{\tablefoottext{a}{This line was mapped in OTF mode.}}

\end{table*}

AFGL 2591 is a well studied high-mass star-forming region at a distance of 3.3 kpc \citep{Rygl12}.
The source is one of the rare cases of a massive star-forming region in relative isolation so that we can study physical parameters
like density, temperature, and velocity structure without confusion from other nearby objects even with single-dish telescopes.
Large amounts of gas and dust toward this source block our view at optical wavelengths,
but result in bright infrared emission. This source is also associated with a weak radio continuum source and with a bipolar outflow.
There is very high velocity CO mid-infrared absorption \citep{Mitchell89}. The luminosity of this source is about 2.0$\times$10$^5$ L$_{\sun}$ \citep{Sanna12}.

\begin{figure*}
    \centering
    \includegraphics[angle=0, width=\textwidth]{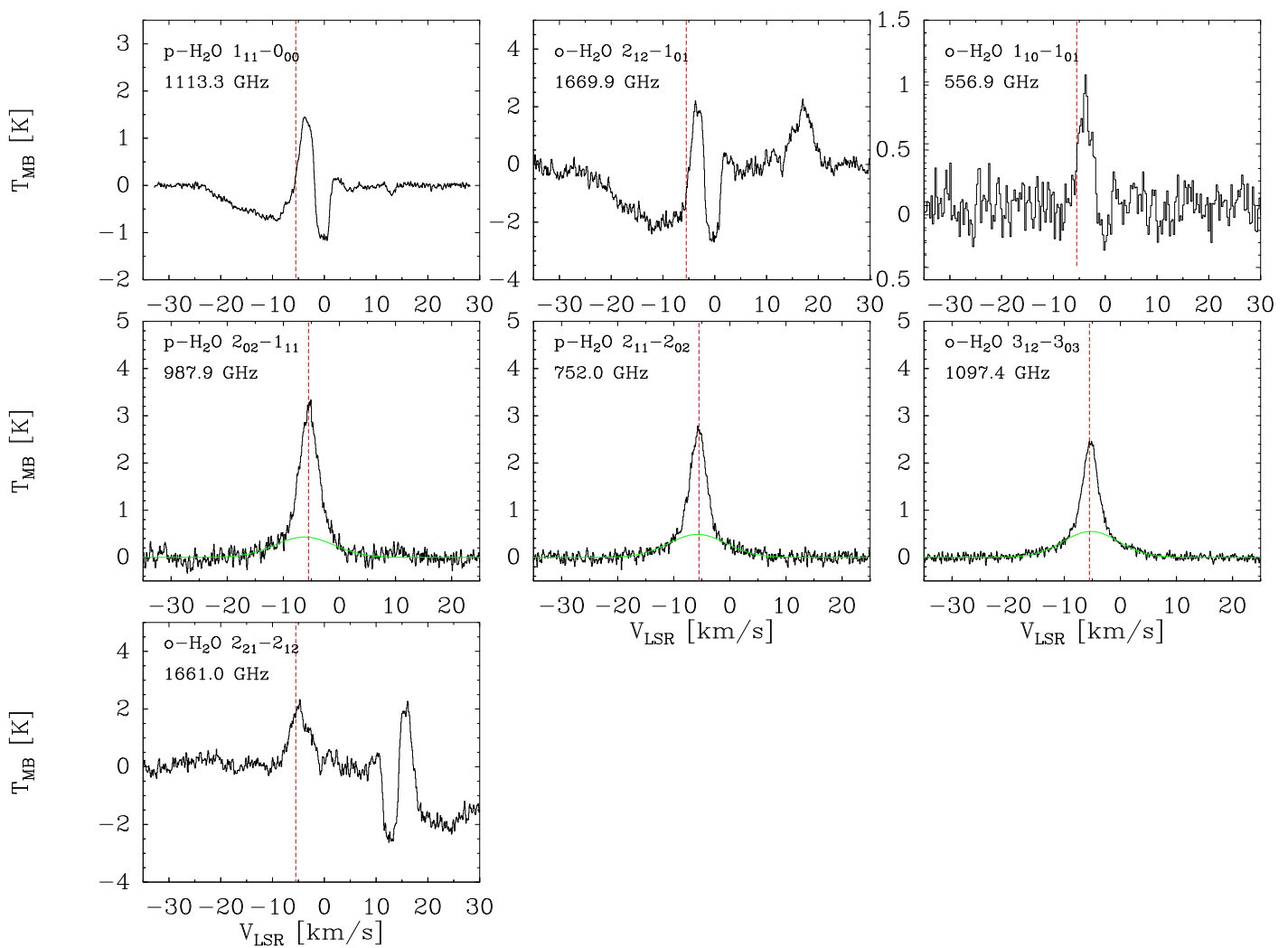}
       \caption{Spectra of H$_2$$^{16}$O lines toward the high-mass star-forming region AFGL 2591. Dashed red lines are drawn at
       the source velocity $V_{\rm LSR}$=$-$5.5 km s$^{-1}$. Green lines present the outflow components. We note that the H$_2$O $2_{12}-1_{01}$ and $2_{21}-2_{12}$ lines occur in opposite sidebands of the same spectrum, causing the ``features'' near 20 km s$^{-1}$.}
          \label{Fig:1}
  \end{figure*}

\begin{figure*}
    \centering
    \includegraphics[angle=0, width=\textwidth]{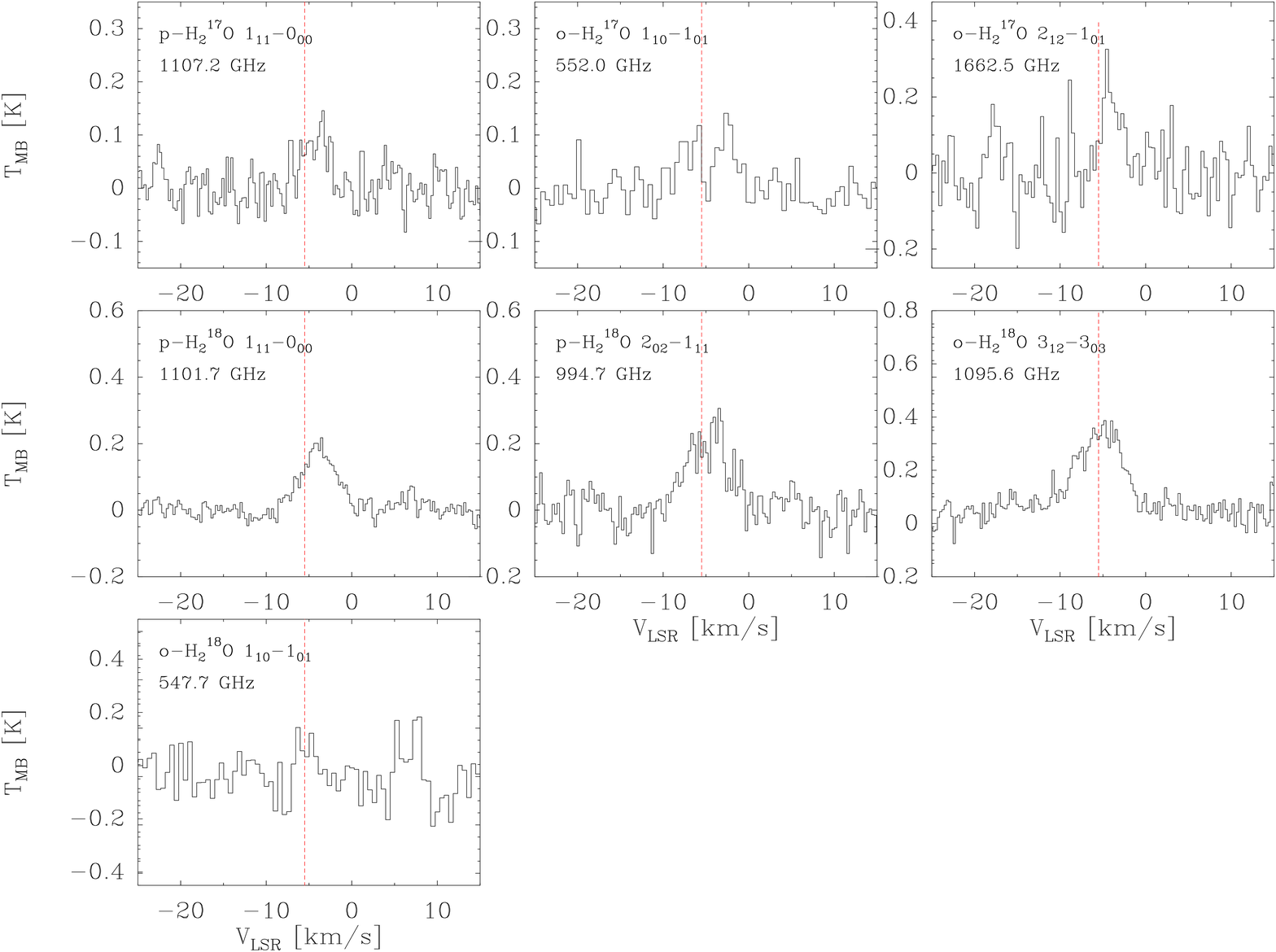}
       \caption{Spectra of H$_2$$^{17}$O and H$_2$$^{18}$O lines toward the high-mass star-forming region AFGL 2591. Dashed lines are drawn at the source velocity $V_{\rm LSR}$=$-$5.5 km s$^{-1}$.}
        \label{Fig:2}
  \end{figure*}

AFGL 2591 has been observed in water lines over a range of excitation conditions. \citet{Helmich96} found more than 30 lines within the bending vibration of water
between 5.5 and 6.6 $\mu$m using ISO-SWS. No lines of H$_2$$^{18}$O or vibrationally-excited H$_2$O were seen.
The ISO data indicate a high excitation temperature ($>$200 K) and a high abundance ($\sim$2$-$6$\times$10$^{-5}$).
\citet{Boonman03} observed gas-phase H$_2$O lines between 5 and 540 $\mu$m with ISO and SWAS.
They found that ice evaporation in the warm inner envelope and freeze-out in the cold outer part together with pure gas-phase chemistry reproduces the H$_2$O observations.
However, these conclusions were based on spectrally unresolved data that could not separate envelope and outflow.
\citet{vdTak06} and \citet{Wang12} presented ground-based observations of H$_2$$^{18}$O toward AFGL 2591
with the Plateau de Bure Interferometer, which have high resolution but are limited to a single line.
These data confirm the abundance ``jump'' inferred from the ISO/SWAS data, and suggest the presence of a circumstellar disk.

This paper uses {\it Herschel}/HIFI observations of water lines toward AFGL 2591 to learn about physical processes in this region and measure the abundance of water
in its various physical components as a step towards understanding the process of high-mass star formation. With its much higher spatial and spectral
resolution and higher sensitivity than previous space missions, {\it Herschel}-HIFI is able to resolve the line profiles and detect isotopic lines, providing essential information
on the physical and chemical structure of the region.

Our observations of AFGL 2591 are summarized in Sect. 2.
In Sect. 3 we show the observational results and simple analysis. Section 4 presents the analysis of
physical conditions using observational data and radiative transfer modeling.
Finally, we discuss our results in Sect. 5.


\section{Observations}
   AFGL 2591 was observed with the Heterodyne Instrument for the Far-Infrared
   \citep[HIFI;][]{deGraauw10} onboard ESA's {\it Herschel} Space Observatory \citep{Pilbratt10}.
   These observations were conducted between March and June 2010,
   using the dual beam switch (DBS) mode as part of
   the guaranteed time key program Water In Star-forming regions with {\it Herschel} \citep[WISH;][]{vDishoeck11}.
   The coordinates of the observed position in AFGL 2591 are 20$^{\rm h}$29$^{\rm m}$24$^{\rm s}$.87
   and +40$\degr$11$\arcmin$19.5$\arcsec$ (J2000).

  Data were taken simultaneously in horizontal and vertical polarizations using both the correlator-based high-resolution spectrometer (HRS) and the acousto-optical wide-band spectrometer (WBS) with a 1.1 MHz resolution.
  We used the double beam switch observing mode with a throw of 3$\arcmin$.
  HIFI receivers are double sideband with a sideband ratio close to unity.
   Currently, the flux scale accuracy is estimated to be about 10\% for bands 1 and 2, 15\% for bands
   3 and 4, and 20\% in bands 6 and 7 \citep{Roelfsema12}.
We show the HRS spectra in Fig. 1 and 2, with the exception of the p-H$_2$O $1_{11}-0_{00}$, o-H$_2$O $2_{12}-1_{01}$,
o-H$_2$O $2_{21}-2_{12}$, and o-H$_2$$^{17}$O $2_{12}-1_{01}$ lines,
for which WBS spectra were used since the velocity range covered by the HRS was insufficient.

AFGL 2591 was also mapped with HIFI in OTF mode in the o-H$_2$O $1_{10}-1_{01}$, p-H$_2$O $1_{11}-0_{00}$, and p-H$_2$O $2_{02}-1_{11}$ lines.
These observations were carried out between November and December 2010.
We have taken the o-H$_2$O $1_{10}-1_{01}$ and o-H$_2$$^{18}$O $1_{10}-1_{01}$ lines from the central positions of the maps
since we do not have data for these two lines using the double beam switch observing mode.
A full analysis of these maps will be presented elsewhere.

   The frequencies, energy of the upper levels, system temperatures, integration times,
   the beam size and efficiency, rms noise level at a given spectral resolution for each of the lines are provided in Table 1.
   The calibration of the data was performed in the {\it Herschel} interactive processing environment
   \citep[HIPE;][]{Ott10} version 8.0. The resulting Level 2 double sideband (DSB) spectra were exported
   to the FITS format for a subsequent data reduction and analysis using
   the IRAM GILDAS\footnote{http://www.iram.fr/IRAMFR/GILDAS/} package.
   These lines are not expected to be polarized, thus, after inspection, data from the two polarizations were averaged together.


\section{Results}

The HIFI spectra of AFGL 2591 show strong emission and absorption by H$_2$O (Fig. 1), and weaker emission in H$_2$$^{18}$O and H$_2$$^{17}$O lines (Fig. 2).
The line profiles differ considerably between the ground-state levels of H$_2$O, its excited levels, and its isotopologs.

The ground-state lines of the main isotopologue (p-H$_2$O $1_{11}-0_{00}$, o-H$_2$O $2_{12}-1_{01}$, o-H$_2$O $1_{10}-1_{01}$) show a mix of emission and absorption,
as found before for DR21 (van der Tak et al 2010) and W3 IRS5 (Chavarr\'{\i}a et al 2010).
First, all three lines show an emission feature of the source at $V_{\rm LSR}$=$-$3 km s$^{-1}$,
somewhat red-shifted with respect to the $V_{\rm LSR}$ of the source at $V_{\rm LSR}$=$-$5.5 km s$^{-1}$ \citep{vdTak99}.
The emission feature seems to be related to an expansion of the outer envelope.
   The expansion is probably powered by outflows which are known to exist in AFGL 2591 \citep{Lada84}.
   The blue side of the emission is smooth because of the outflow, but the red side is sharply truncated because of absorption by a foreground cloud at $\sim$0 km s$^{-1}$.

Second, a broad and asymmetric absorption component occurs near $V_{\rm LSR}$=$-$10 km s$^{-1}$ which from its shape has a likely origin in a wind.
This broad absorption component is only detected in the p-H$_2$O $1_{11}-0_{00}$ and o-H$_2$O $2_{12}-1_{01}$ lines.
We probably do not see it in the o-H$_2$O $1_{10}-1_{01}$ line because the signal-to-noise ratio on the continuum is not high enough.

Third, the ground-state H$_2$O line profiles show evidence for two foreground clouds.
The narrow absorption component around $V_{\rm LSR}$=0 km s$^{-1}$ is detected in all three lines,
and corresponds to a cloud known from ground-based observations \citep{vdTak99}.
A second, weaker absorption feature near $V_{\rm LSR}$=13 km s$^{-1}$ is only seen in the p-H$_2$O $1_{11}-0_{00}$ line.
The continuum at o-H$_2$O $1_{10}-1_{01}$ line is presumably too weak to see this feature, and the o-H$_2$O $2_{12}-1_{01}$ spectrum is blended
with the o-H$_2$O $2_{21}-2_{12}$ line. This cloud is not known from ground-based observations,
and is also seen in HF observations \citep{Emprechtinger12}
but not in CO observations by {\it Herschel}/HIFI of this source \citep{vdWiel13},
so it is probably a diffuse cloud.

The excited-state lines of H$_2$O (p-H$_2$O $2_{02}-1_{11}$, p-H$_2$O $2_{11}-2_{02}$,
   and o-H$_2$O $3_{12}-3_{03}$) appear purely in emission and show two velocity components.
   The components have Gaussian shapes, one being wider (FWHM=11$-$12 km s$^{-1}$)
   than the other (FWHM=3$-$4 km s$^{-1}$).
   Studies of low-mass protostars from the WISH program \citep{Kristensen10, Kristensen12}
   refer to these components as the ``broad'' and ``narrow'' components, but we will call them the ``outflow'' and ``envelope'' components, after their likely physical origin.
We assume that the broad component is due to the high-velocity outflow
   associated with the protostar seen in absorption in p-H$_2$O $1_{11}-0_{00}$ and o-H$_2$O $2_{12}-1_{01}$ lines,
   even though it covers a somewhat smaller velocity range.
   The narrow component is potentially associated with the protostellar envelope.

The lines of H$_2$$^{18}$O and H$_2$$^{17}$O appear purely in emission and are dominated by the envelope, centered around $V_{\rm LSR}$=$-$5 km s$^{-1}$.
In addition, the H$_2$$^{18}$O lines around 1~THz may show a broadening due to a weak outflow component.
Likewise, the high-frequency o-H$_2$$^{17}$O $1_{10}-1_{01}$ line appears to be broader than the other two H$_2$$^{17}$O lines.

We extracted line parameters from the observed profiles by fitting Gaussians;
Table 2 gives the results for the emission lines and Table 3 for the absorption lines. For the emission features,
we fitted the p-H$_2$O $2_{02}-1_{11}$, p-H$_2$O $2_{11}-2_{02}$,
   and o-H$_2$O $3_{12}-3_{03}$ line profiles assuming two velocity components, while the other emission lines are
   fitted as one velocity component. The H$_2$$^{18}$O lines appear broader than the narrow emission components seen in H$_2$O (see Table 2),
so that the profiles are probably a mixture of envelope and outflow emission seen at limited signal-to-noise ratio.
The o-H$_2$$^{17}$O $1_{10}-1_{01}$ might show two components but the signal-to-noise ratio is not good enough
so that we assume that this line has one component.

\begin{table*}
\caption{Parameters of the H$_2$O, H$_2$$^{18}$O, and H$_2$$^{17}$O emission line profiles obtained from Gaussian fits.}
\label{table:2}
\centering
\begin{tabular}{l l c c c c}
\hline\hline
Molecule & Transition & $\int T_{\rm MB}dV$ & $V_{\rm LSR}$ & $\Delta V$    & $T_{\rm MB}$ \\
         &            & (K km s$^{-1}$)     & (km s$^{-1}$) & (km s$^{-1}$) & (K)          \\
\hline
p-H$_2$O        & $2_{02}-1_{11}$ & 6.1 (0.5) & -6.3 (0.3) & 10.1 (0.6) & 0.6 \\
                &                 & 9.9 (0.5) & -5.4 (0.1) &  3.7 (0.1) & 2.5 \\
p-H$_2$O        & $2_{11}-2_{02}$ & 5.9 (0.2) & -5.8 (0.1) & 11.1 (0.5) & 0.5 \\
                &                 & 7.5 (0.2) & -5.6 (0.1) &  3.3 (0.1) & 2.1 \\
o-H$_2$O        & $3_{12}-3_{03}$ & 6.4 (0.1) & -5.3 (0.1) & 10.6 (0.2) & 0.6 \\
                &                 & 5.9 (0.1) & -5.3 (0.1) &  3.1 (0.0) & 1.8 \\
\hline
p-H$_2$$^{18}$O & $1_{11}-0_{00}$ & 1.1 (0.1) & -4.0 (0.1) & 5.0 (0.2) & 0.2 \\
p-H$_2$$^{18}$O & $2_{02}-1_{11}$ & 1.3 (0.1) & -4.8 (0.2) & 5.5 (0.4) & 0.2 \\
o-H$_2$$^{18}$O & $3_{12}-3_{03}$ & 1.3 (0.1) & -5.2 (0.1) & 5.6 (0.2) & 0.2 \\
\hline
p-H$_2$$^{17}$O & $1_{11}-0_{00}$ & 0.4 (0.1) & -4.3 (0.2) & 3.8 (0.4) & 0.1 \\
o-H$_2$$^{17}$O & $1_{10}-1_{01}$ & 0.5 (0.1) & -4.2 (0.6) & 7.7 (1.2) & 0.1 \\
o-H$_2$$^{17}$O & $2_{12}-1_{01}$ & 0.5 (0.1) & -4.2 (0.2) & 2.3 (0.5) & 0.2 \\
\hline
\end{tabular}
\end{table*}

\begin{table*}
\caption{Column densities estimated from p-H$_2$O $1_{11}-0_{00}$, o-H$_2$O $2_{12}-1_{01}$ and o-H$_2$O $1_{10}-1_{01}$ absorption line profiles.}
\label{table:3}
\centering
\begin{tabular}{l c l c c c}
\hline\hline
Line & $T_{\rm cont}$     & vel. range & $\tau$\tablefootmark{a} & $N$  \\
     &  (K)&    (km s$^{-1}$)     &                              & (10$^{12}$ cm$^{-2}$) \\
\hline
p-H$_2$O $1_{11}-0_{00}$ & 1.1$\pm$0.2 & -24 to -5.5  & 0.5$\pm$0.1 & 20.9$\pm$4.4     \\
                         &             & -2  to 1.5   & 1.6$\pm$0.4 & 13.3$\pm$2.8     \\
                         &             & 12  to 14    & 0.1$\pm$0.1 &  0.4$\pm$0.1      \\
\hline
o-H$_2$O $2_{12}-1_{01}$ & 2.3$\pm$0.5 & -24 to -5.5  & 0.9$\pm$0.2 & 73.8$\pm$20.9 \\
                         &             & -2  to 1.5   & 1.8$\pm$0.5 & 38.9$\pm$12.6 \\
                         &             & 12  to 14    & $-$         & $-$ \\
\hline
o-H$_2$O $1_{10}-1_{01}$ &0.1$\pm$0.1 & -24 to -5.5   & $-$         & $-$ \\
                         &            & -2  to 1.5    & 0.5$\pm$0.1 & 25.6$\pm$3.6 \\
                         &            & 12  to 14     & $-$         & $-$ \\
\hline
\end{tabular}
\tablefoot{\tablefoottext{a}{$\tau$ is the velocity-averaged optical depth}}
\end{table*}


\section{Analysis}
\subsection{Absorption components}
The ground-state lines of the main isotopologue (p-H$_2$O $1_{11}-0_{00}$, o-H$_2$O $2_{12}-1_{01}$, o-H$_2$O $1_{10}-1_{01}$)
show three absorption components:
1) the broad $V_{\rm LSR}$=$-$10 km s$^{-1}$ component due to the molecular outflow, 2) the narrow $V_{\rm LSR}$= 0 km s$^{-1}$ component
due to the known foreground cloud, and 3) the narrow $V_{\rm LSR}$= 13 km s$^{-1}$ component due to a new diffuse foreground cloud.
We derived the optical depth in these three components using the expression
\begin{equation}
\tau=-{\rm ln}\left(\frac{T_{\rm line}}{T_{\rm cont}}\right) \,,
\end{equation}
where $T_{\rm cont}$ is the single side band (SSB) continuum intensity. This expression
assumes that the sideband gain ratio is unity and that the continuum is completely covered by the absorbing layer.
We applied a linear baseline fit in the vicinity of the absorption line to derive the continuum intensity at the absorption peak.
Deriving the optical depth from the line-to-continuum ratio is based on the assumption that
the excitation temperature is negligible with respect to the continuum temperature.

In the following analysis we assume that all water molecules are in the ortho and para ground states, so that
the velocity integrated absorption is related to the molecular column density by
   \begin{equation}
      N = \frac{8\pi\nu^{3}g_l}{c^{3}Ag_u}\int{\tau dV} \,,
   \end{equation}
   where $N$ is the column density, $\nu$ the frequency, $c$ the speed of light, and $\tau$ is the optical depth.
   $A$ stands for the Einstein-A coefficient and $g_l$ and $g_u$ are the degeneracy of the lower and the upper
   level of the transition.
   Subsequently, we integrated over the velocity ranges given in Table 3 to determine the column
density for each component. The derived column densities and the optical depth
of the three components are listed also in Table 3.
For the velocity range from -24 km s$^{-1}$ to -5.5 km s$^{-1}$, the column density is $\sim$2$-$7 10$^{13}$ cm$^{-2}$ and the optical depth is $\sim$0.5$-$1. On the other hand, the column density is $\sim$1$-$4 10$^{13}$ cm$^{-2}$ and the averaged optical depth is $\sim$1.6$-$2 for the velocity range from -2 km s$^{-1}$ to 1.5 km s$^{-1}$.

More information about the physical conditions in the foreground clouds comes from the ortho-to-para ratio of H$_2$O.
If this ratio is thermalized, it should rise from $\sim$1 at low temperatures ($\sim$15 K) to $\sim$3 at high temperatures ($>$40 K) as shown by \citet{Mumma87}. Since we have data for ortho- and para-H$_2$O, we can determine the o/p ratio,
although the dynamic range of the absorption data is limited by the signal-to-noise ratio on one hand and by saturation on the other.

For narrow component, we determine the ortho/para ratio using the ground state of the o-H$_2$O $1_{10}-1_{01}$ and the p-H$_2$O $1_{11}-0_{00}$ lines and we find a lower o/p ratio of $\sim$1.9$\pm$0.4 in the narrow component, suggesting a lower temperature for the foreground cloud.
On the other hand, for the broad component we do not see it in the o-H$_2$O $1_{10}-1_{01}$ line so we used the second ground state ortho-H$_2$O line, o-H$_2$O $2_{12}-1_{01}$ transition at 1670 GHz. The o/p ratio of the broad component is around three ($\sim$3.5$\pm$1.0),
which is reasonable because the gas in the outflow is probably warm as it is heated by shocks.

Similar results have been found for diffuse absorbing clouds toward continuum sources in the Galactic plane \citep{Flagey13}.
Their presented water o/p ratios were consistent with the high-temperature limit value of 3, with lower values for the clouds with the highest column densities. Presumably interstellar UV radiation does not fully penetrate those clouds, so that photo-electric heating of the gas is less efficient.
Indeed, the total (ortho + para) H$_2$O column densities of the foreground clouds of AFGL 2591 are as high as $\sim$ 5$\times$10$^{13}$ cm$^{-2}$ assuming an ortho/para ratio of 3, similar to the values found for the clouds toward NGC 6334 I, which also have a similar ortho/para ratio of H$_2$O \citep{Emprechtinger10}.


\subsection{Emission components}

   To estimate the column densities and rotation temperatures of water in the envelope and outflow of AFGL 2591,
   we construct rotation diagrams for the H$_2$$^{17}$O and H$_2$$^{18}$O lines for the envelope
   and for the broad component of the p-H$_2$O $2_{02}-1_{11}$, p-H$_2$O $2_{11}-2_{02}$,
   and o-H$_2$O $3_{12}-3_{03}$ lines for the outflow.
   We assume an ${}^{16}$O/${}^{18}$O ratio of 550 and an ${}^{18}$O/${}^{17}$O ratio of 4 \citep{Wilson94},
   and that the H$_2$$^{17}$O and H$_2$$^{18}$O lines have the same excitation temperature.

First, we assume that (1) the lines are
optically thin, (2) the emission fills the telescope beam, and (3) all level populations can be characterized by a single excitation temperature $T_{\rm rot}$ and use following equations.
The column densities of the molecules in the upper level $N_{\rm u}$ are related to the measured integrated intensities,
   $\int{T_{\rm mb}d\rm V}$ \citep{Linke79, Blake84, Blake87, Helmich94} by

   \begin{equation}
      N_u/g_u = \frac{N_{\rm tot}}{Q(T_{\rm rot})}e^{-E_u/T_{\rm rot}}=\frac{1.67 \cdot 10^{14}}{\nu\mu^2S}\int{T_{\rm mb}dV} \,,
   \end{equation}
where $\mu$ is the permanent dipole moment,
  $N_{\rm tot}$ is the total column density, $Q(T_{\rm rot})$ is the partition function for the rotation temperature
  $T_{\rm rot}$, and $S$ is the line strength value.
  Thus, a logarithmic plot of the quantity on the right-hand side of equation as a function of $E_u$ provides a straight line
with slope $1/T_{\rm rot}$ and intercept $N_{\rm tot}/Q(T_{\rm rot})$.

\begin{figure}
   \centering
   \includegraphics[angle=-90, width=8cm]{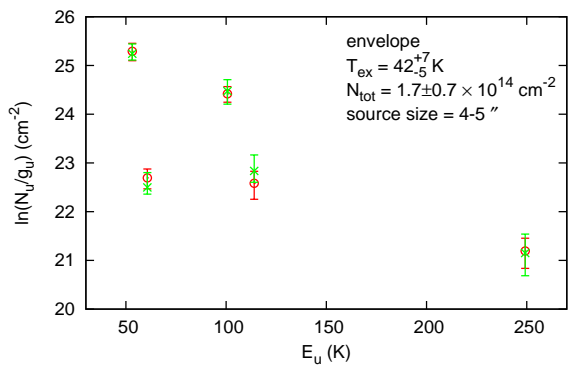}
      \caption{Rotation diagram for the H$_2$$^{18}$O and H$_2$$^{17}$O emission components (envelope). Red open circles are the data observed with
{\it Herschel}/HIFI. The green crosses represent the best-fit model from the the population diagram analysis.
}
         \label{Fig:3}
   \end{figure}

\begin{figure}
   \centering
   \includegraphics[angle=-90,width=8cm]{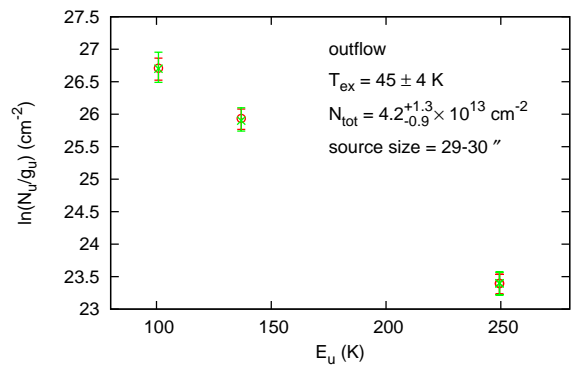}
      \caption{Rotation diagram for the broad emission components seen in p-H$_2$O $2_{02}-1_{11}$,
                               p-H$_2$O $2_{11}-2_{02}$, and o-H$_2$O $3_{12}-3_{03}$ (outflow). Red open circles are the data observed with
{\it Herschel}/HIFI. The green crosses represent the best-fit model from the population diagram analysis.
}
         \label{Fig:4}
   \end{figure}

To test our assumption of low optical depth in the above analysis, we carry out a simple estimate of the line optical depths
comparing the observed H$_2$$^{16}$O-to-H$_2$$^{18}$O and H$_2$$^{18}$O-to-H$_2$$^{17}$O line ratios to the isotopic ratios,
based on the assumption that the excitation temperatures of the corresponding transitions is the same across isotopologues.
The measured peak intensity ratios of the p-H$_2$$^{16}$O $2_{02}-1_{11}$-to-p-H$_2$$^{18}$O $2_{02}-1_{11}$
and o-H$_2$$^{16}$O $3_{12}-3_{03}$-to-p-H$_2$$^{18}$O $3_{12}-3_{03}$ lines are 14 and 10, respectively,
which is well below the isotopic ratio and indicates an optical depth of 3.6$-$3.9 for H$_2$$^{16}$O.
In contrast, our observed p-H$_2$$^{18}$O $1_{11}-0_{00}$-to-p-H$_2$$^{17}$O $1_{11}-0_{00}$ line ratio is $\sim$2,
close to the isotopic ratio, which means that the optical depth of the H$_2$$^{18}$O lines is $\la$1.

We now construct rotation diagrams which take into account the effect of optical depth.
We define the optical depth correction factor $C_\tau$ (= ${\tau}/({1-e^{-\tau}})$).
If the source does not fill the beam, then the correct upper level column density is greater than
that obtained assuming the beam to be filled by a factor equal to the beam dilution $f (= \Delta\Omega_a / \Delta\Omega_s$),
with $\Omega_s$ the size of the emission region and $\Omega_a$ the size of the telescope beam.
The equation mentioned for the rotation diagram method as Eq. 3 can be modified
to include the effect of optical depth $\tau$ through the factor
$C_\tau$ and beam dilution $f$ \citep{Goldsmith99}:

\begin{equation}
     {\rm ln}\left(\frac{N_u}{g_u}\right) = {\rm ln}\left(\frac{N_{\rm tot,thin}}{Q(T_{\rm rot})}\right) - \frac{E_{\rm u}}{kT_{\rm ex}} - {\rm ln}(C_\tau) + {\rm ln}(f).
\end{equation}

According to Eq. 4, for a given upper level, $N_u$ can be evaluated
from a set of $N_{\rm tot,thin}$, $T_{\rm ex}$, $f$ and $C_\tau$. Since $C_\tau$ is a function
of $N_{\rm tot,thin}$ and $T_{\rm ex}$, the independent parameters are therefore $N_{\rm tot,thin}$,
$T_{\rm ex}$ and $f$, for which we solve self-consistently.
A $\chi^2$ minimization gives best-fit values of $N_{\rm tot,thin}$, $T_{\rm ex}$, and source sizes.
In Fig. 3 \& 4, rotation diagrams for water molecules in AFGL 2591 are presented. Red open circles are the data observed with
{\it Herschel}/HIFI and the green crosses represent the best-fit model from population diagram analysis (using Eq. 4.).

Figure 3 shows the rotation diagram for H$_2$$^{18}$O and H$_2$$^{17}$O, which is associated with the envelope.
We construct rotation diagram for the envelope using five H$_2$$^{18}$O and H$_2$$^{17}$O emission lines because we do not detect the o-H$_2$$^{18}$O $1_{10}-1_{01}$ line and p-H$_2$$^{18}$O $1_{11}-0_{00}$ line and p-H$_2$$^{17}$O $1_{11}-0_{00}$ line have the same energy of the upper level so we use the p-H$_2$$^{17}$O $1_{11}-0_{00}$ line which is optically thin.
The excitation temperature is estimated to be 42 $\pm$ 7 K with column density
of (1.7 $\pm$ 0.7) $\times$ 10$^{14}$ cm$^{-2}$ distributed over a $\sim$4$\arcsec$ region. Line opacities of all the transitions
used are estimated to be less than 14, which are consistent with those derived from the observed H$_2$$^{16}$O-to-H$_2$$^{18}$O line ratios to the isotopic ratios.

Figure 4 presents the same analysis for the broad components seen in p-H$_2$O $2_{02}-1_{11}$, p-H$_2$O $2_{11}-2_{02}$, and o-H$_2$O $3_{12}-3_{03}$.
The broad component is likely related to the outflow, for which we find an excitation temperature of 45$\pm$4 K,
a column density of (4.2 $\pm$ 1.0) $\times$ 10$^{13}$ cm$^{-2}$, a source size of $\sim$30$\arcsec$ and line opacities of the three H$_2$O transitions (broad components) used of $<$ 0.03.

\begin{figure*}
\centering
\includegraphics[width=16cm]{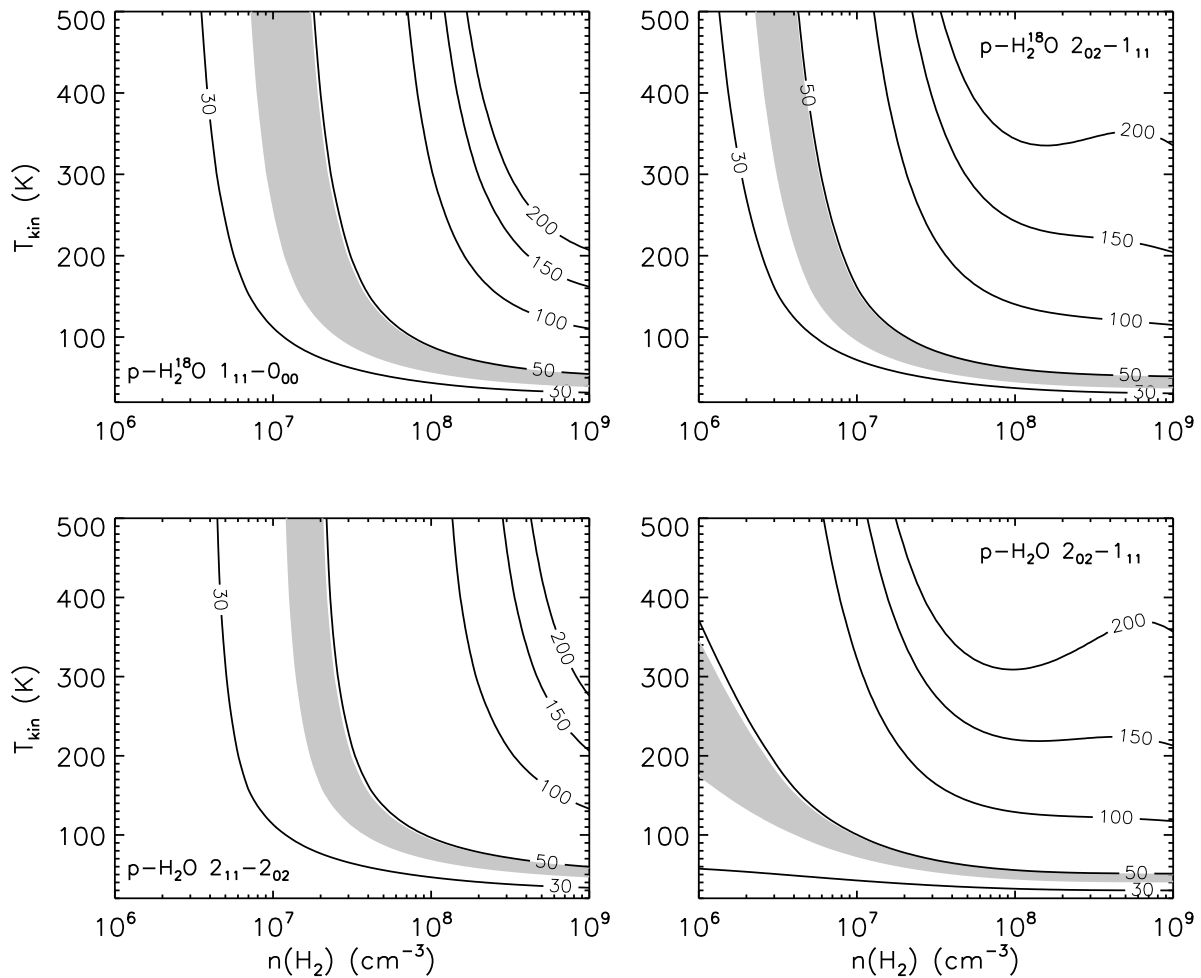}
      \caption{The excitation temperature of p-H$_2$$^{18}$O 1$_{11}-0_{00}$, p-H$_2$$^{18}$O 2$_{02}-1_{11}$ for the envelope (\textit{upper})
               and the p-H$_2$O 2$_{11}-2_{02}$, p-H$_2$O 2$_{02}-1_{11}$ for the outflow (\textit{lower})
               assuming H$_2$O column densities of $N$(H$_2$O) = 1$\times$10$^{14}$ cm$^{-2}$ and $N$(H$_2$O) = 5$\times$10$^{13}$ cm$^{-2}$
               and source sizes of 4$\arcsec$ and 30$\arcsec$ for the envelope and the outflow, respectively
               as a function of kinetic temperature and H$_2$ density calculated with RADEX (Non-LTE, large velocity gradient code).
               The gray areas indicate the derived $T_{\rm ex}$ of envelope and outflow from the LTE rotation diagram analysis.
               }
       \label{Fig:5}
\end{figure*}

\begin{figure*}
\centering
\includegraphics[width=16cm]{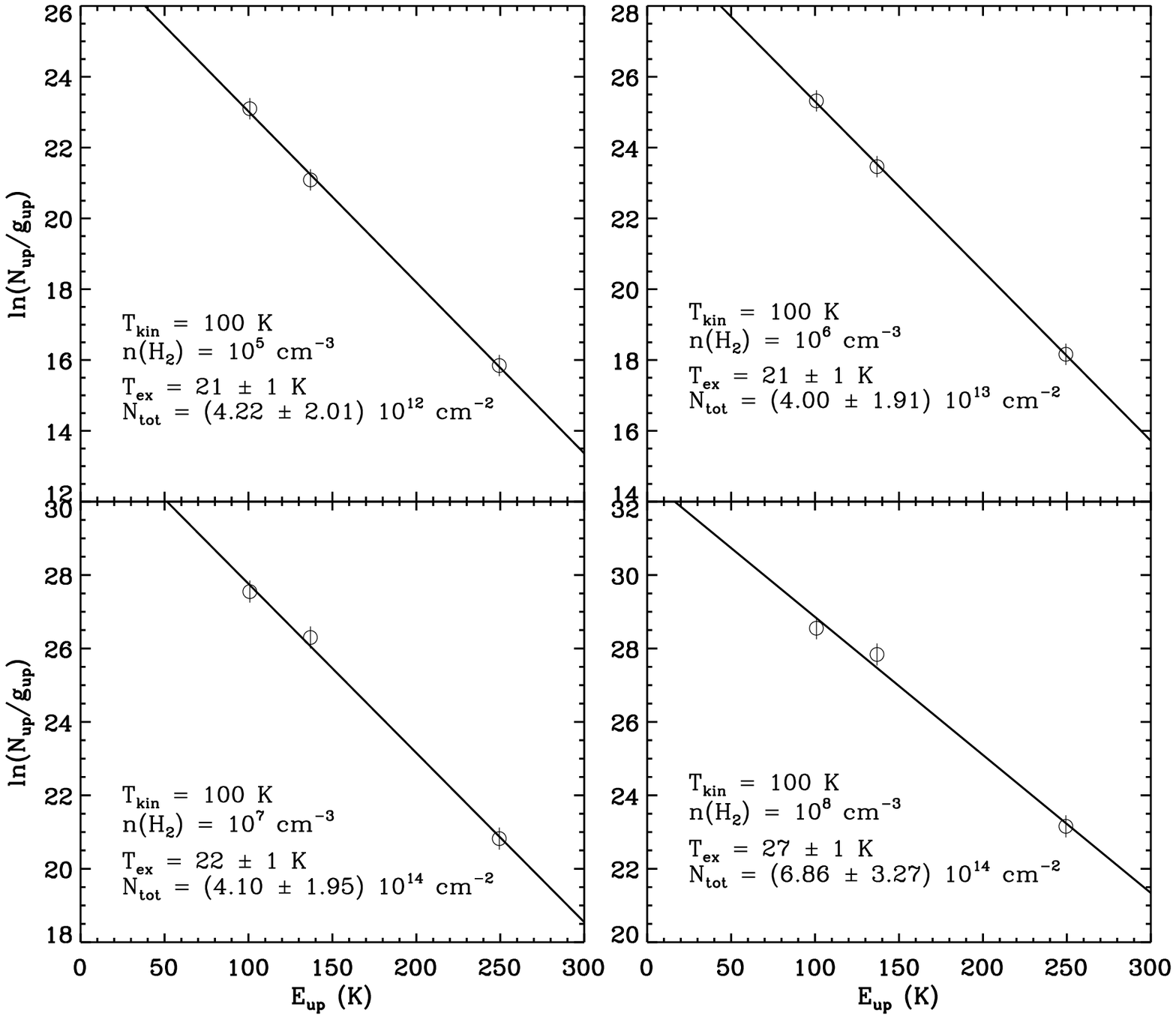}
      \caption{Examples of rotation diagram for the p-H$_2$O $2_{02}-1_{11}$,
                               p-H$_2$O $2_{11}-2_{02}$, and o-H$_2$O $3_{12}-3_{03}$ emission lines originating in the outflow at 
                               a given kinetic temperature of $T_{\rm kin}$ = 100 K from the RADEX calculations (non-LTE analysis). Open circles are the data from the non-LTE models. The overplotted line corresponds to a linear fit to the rotational diagram.}
       \label{Fig:6}
\end{figure*}

\subsection{Non-LTE calculations}

We carried out non-LTE models of H$_2$O using the RADEX code \citep{vdTak07} and state-of-the-art quantum mechanical collision rates of para and ortho H$_2$O with para
and ortho H$_2$ \citep{Daniel11} as provided at the LAMDA database \citep{Schoier05}. The same collision data are used for all isotopologs.
To constrain the H$_2$O excitation, we generated a grid of models with values of $T_{\rm kin}$ between 10 and 1000 K, values of $n$(H$_2$) from 10$^3$ to 10$^9$ cm$^{-3}$,
and fixed the background radiation temperature at 2.73 K.
The line width was fixed at 3 km s$^{-1}$ for the envelope and 10.5 km s$^{-1}$ for the outflow, and we applied a molecular column density of $N$(H$_2$O) = 1$\times$10$^{14}$ cm$^{-2}$ and $N$(H$_2$O) = 5$\times$10$^{13}$ cm$^{-2}$
for the envelope and the outflow, respectively. These molecular column densities are derived by the rotation diagram method (LTE) assuming source sizes of 4$\arcsec$ and 30$\arcsec$.

In the comparison with data, we first focus on the p-H$_2$$^{18}$O 1$_{11}-0_{00}$ and p-H$_2$$^{18}$O 2$_{02}-1_{11}$ lines for the envelope,
which lie close in frequency, so that the effects of beam filling cancel out to first order.
For the outflow, we use the p-H$_2$O 2$_{11}-2_{02}$ and p-H$_2$O 2$_{02}-1_{11}$ lines,
based on the calculations of the p-H$_2$O 2$_{11}-2_{02}$/p-H$_2$O 2$_{02}-1_{11}$ line ratio,
which traces both the gas temperature and density (Fig A.1).

Figure 5 presents the calculated excitation temperatures of the  p-H$_2$$^{18}$O 1$_{11}-0_{00}$ and p-H$_2$$^{18}$O 2$_{02}-1_{11}$ lines for the envelope (\textit{upper})
and of the p-H$_2$O 2$_{11}-2_{02}$ and p-H$_2$O 2$_{02}-1_{11}$ lines for the outflow (\textit{lower}),
as a function of gas density and kinetic temperature.
The calculations show that the derived range of excitation temperatures for the envelope (gray area in upper panels)
based on the rotation diagram method (LTE) indicates a gas density of 7$\times$10$^6-10^8$ cm$^{-3}$  and
a kinetic temperature of $\sim$60$-$230 K indicating subthermal excitation.
On the other hand, the models indicate that
a gas density of $\sim$10$^7-10^8$ cm$^{-3}$ and a kinetic temperature of $\sim$70$-$90 K reproduce the observations of the outflow (gray area in lower panels).

\begin{table*}
\caption{Physical conditions for each components}
\label{table:4}
\centering
\begin{tabular}{l c c c c c c}
\hline\hline
Components & $V_{\rm LSR}$ & FWHM          & $N$(CO)     & $N$(H$_2$)   & $N(\rm {H_2O})$ & $X(\rm {H_2O})$ \\
           & (km s$^{-1}$) & (km s$^{-1}$) & (cm$^{-2}$) & (cm$^{-2}$)  & (cm$^{-2}$)     &                  \\
\hline
Foregound\tablefootmark{a}         & 0    &       & $\ga$ 3 $\times$ 10$^{17}$\tablefootmark{d} & $\ga$ 3 $\times$ 10$^{21}$ &
$\sim$5.2$\times$10$^{13}$\tablefootmark{h}  & $\la$ 1.7 $\times$ 10$^{-8}$ \\
Weaker foreground\tablefootmark{a} & +13  &       & $\la$ 6 $\times$ 10$^{15}$\tablefootmark{d} & $\sim$8$\times$10$^{20}$\tablefootmark{e}
$-$ 6$\times$10$^{21}$ & $\sim$1.6 $\times$ 10$^{12}$\tablefootmark{h}  & $\sim$2.7$\times$ 10$^{-10}$ $-$ 2$\times$10$^{-9}$ \\
Envelope\tablefootmark{b}          & -5.5 & 5--6   & $\sim$10$^{18}$\tablefootmark{d} & $\sim$10$^{22}$ &
$\sim$1.7$\times$10$^{14}$\tablefootmark{i}  & 1.7$\times$ 10$^{-8}$ \\
                                   &      &       & 7.2$\times$10$^{18}$\tablefootmark{f} & 7.2$\times$ 10$^{22}$ & & 2.4$\times$10$^{-9}$ \\
                                   &      &       & 7.29$\times$10$^{18}$\tablefootmark{g} & 2.7$\times$10$^{22}$ & & 6.3$\times$10$^{-9}$  \\
Outflow\tablefootmark{c}           & -10  & 10--11 & 1.5$\times$ 10$^{18}$\tablefootmark{d} & 1.5$\times$ 10$^{22}$ &
$\sim$4.2$\times$10$^{13}$\tablefootmark{j} & 2.8$\times$10$^{-9}$     \\
                                   &      &       & 6.6$\times$10$^{18}$\tablefootmark{f} & 6.6$\times$10$^{22}$ & &
6.4$\times$10$^{-10}$   \\                                    \\
\hline
\end{tabular}
\tablefoot{
           \tablefoottext{a}{$V_{\rm LSR}$ of two foreground components are from the p-H$_2$O $1_{11}-0_{00}$ line.} \\
           \tablefoottext{b}{$V_{\rm LSR}$ and FWHM of the envelope component is from H$_2$$^{18}$O lines.} \\
           \tablefoottext{c}{$V_{\rm LSR}$ and FWHM of the outflow component is from broad components seen the p-H$_2$O $2_{02}-1_{11}$,
                               p-H$_2$O $2_{11}-2_{02}$, and o-H$_2$O $3_{12}-3_{03}$ lines.} \\
           \tablefoottext{d}{\citet{vdWiel13}} \\
           \tablefoottext{e}{A lower limit from HF 1--0 observations. \citet{Emprechtinger12}} \\
           \tablefoottext{f}{\citet{Mitchell89}}\\
           \tablefoottext{g}{CO column density from C$^{18}$O $J$=9--8 observations assuming an abundance of $^{16}$O/$^{18}$O of 540. \citet{SanJG13}} \\
           \tablefoottext{h}{The total (ortho+para) H$_2$O column density assuming an ortho/para ratio of 3.} \\
           \tablefoottext{i}{H$_2$O column density for the envelope for a source size of 4$\arcsec$.} \\
           \tablefoottext{j}{H$_2$O column density for the outflow for a source size of 30$\arcsec$.}
           }
\end{table*}

To compare LTE and non-LTE calculations, we construct the rotation diagrams for the broad components of the p-H$_2$O $2_{02}-1_{11}$, p-H$_2$O $2_{11}-2_{02}$, and o-H$_2$O $3_{12}-3_{03}$ lines using the RADEX calculation as data points at different kinetic temperatures, $T_{\rm kin}$, and densities, $n(\rm {H}_2)$. We apply a molecular column density of $N$(H$_2$O) = 5$\times$10$^{13}$ cm$^{-2}$ and a line width of 10.5 km s$^{-1}$.
Figure 6 presents the rotation diagrams for the broad components of the three excited-state lines of H$_2$O assuming a kinetic temperature of $T_{\rm kin}$ = 100 K as examples. Open circles are the data from the RADEX calculations. The overplotted line represents a linear fit to the rotational diagram.
The rotation temperatures, $T_{\rm rot}$, are below the input kinetic temperature of $T_{\rm kin}$ = 100 K.
At higher H$_2$ density the rotation temperature is higher than at low H$_2$ density 
but our results show that the rotation temperature is well below the kinetic temperature, even at
$n(\rm {H}_2)$ = 10$^8$ cm$^{-3}$.
Higher densities are implausible for AFGL 2591 at the spatial scales probed by our data.

\section{Discussion}

We estimate the H$_2$O abundance in the various physical components of AFGL 2591 (Table 4) 
assuming a constant abundance for the protostellar envelope.
First, \citet{vdWiel13} presents an absorption feature seen in CO 5--4 and JCMT data of CO 3--2 near 0 km s$^{-1}$, which is known to be a foreground component. They derived a column density $N$(H$_2$) $\ga$ 3$\times$10$^{21}$ cm$^{-2}$ assuming an abundance of CO/H$_2$ of 10$^{-4}$.
Assuming an ortho/para ratio of 3, we estimate the total (ortho+para) H$_2$O column density for the 0 km s$^{-1}$ foreground component using the p-H$_2$O $1_{11}-0_{00}$ line. We find that the total H$_2$O column density is $\sim$ 5.2$\times$10$^{13}$ cm$^{-2}$.
and the abundance of H$_2$O is $\la$ 1.7$\times$10$^{-8}$, consistent with ion-molecule chemistry.
In addition to the 0 km s$^{-1}$ foreground component, there is another weak absorption component at 13 km s$^{-1}$, for which
\citet{vdWiel13} obtained limits on the H$_2$ column density.
As an upper limit, they estimated $N$(H$_2$) $\la$ 6$\times$10$^{21}$ cm$^{-2}$ based on their $^{13}$CO 1--0 spectrum.
In addition, they used the HF 1--0 spectrum presented by \citet{Emprechtinger12} to obtain a lower limit for the column density of the 13 km s$^{-1}$ foreground component. They found that $N$(H$_2$) $\ga$ 8$\times$10$^{20}$ cm$^{-2}$ assuming HF/H$_2$ = 3.6$\times$10$^{-8}$.
We estimate the H$_2$O column density of $N$(H$_2$O) = 1.6$\times$10$^{12}$ cm$^{-2}$ using the p-H$_2$O $1_{11}-0_{00}$ line under the assumptions that an ortho/para ratio is 3,
so that the H$_2$O abundance in this component is $\sim$10$^{-9}$. This lower abundance compared with the 0~km s$^{-1}$ component is consistent with enhanced photodissociation in diffuse gas.

In order to derive the H$_2$O abundance in the outflow and envelope components, we use two methods to estimate the H$_2$ column density. We adopt the H$_2$O column densities of $\sim$1.7$\times$10$^{14}$ cm$^{-2}$
for the envelope and $\sim$4.2$\times$10$^{13}$ cm$^{-2}$ for the outflow, in source sizes of 4$\arcsec$ and 30$\arcsec$, respectively, based on LTE calculations.
\citet{Mitchell89} detected CO and $^{13}$CO rovibrational absorption lines near 4.7 $\mu$m and found a cold ($\sim$38 K)
and a hot ($\sim$1000 K) component in the quiescent gas centered around --5 km s$^{-1}$, and a blueshifted warm ($\sim$200 K) component.
The cold component is likely the outer envelope of AFGL 2591, while the hot gas is near the infrared source and
is heated by its luminosity. Despite their different temperatures, the CO column densities of the three components are all in the range 5$-$7 $\times$ 10$^{18}$ cm$^{-2}$.
We use the column density from the cold gas for the envelope, and that of the blueshifted
warm gas for the outflow, and derive the column density of H$_2$ using a ratio $N$(${}^{12}$CO)/$N$(H$_2$) = 10$^{-4}$.
We find that the abundance of H$_2$O is 2.4 $\times$ 10$^{-9}$ for the envelope and 6.4 $\times$ 10$^{-10}$ for the outflow.

As a second method, we estimate $N$(H$_2$) from submillimeter observations.
\citet{SanJG13} presented {\it Herschel}/HIFI observations of high-$J$ CO and isotopologues from low-to high-mass star-forming regions.
They estimated a H$_2$ column density for AFGL 2591 of 2.7 $\times$ 10$^{22}$ cm$^{-2}$ in a $\sim$20$\arcsec$ beam using C$^{18}$O $J$=9--8 for an excitation temperature of 75 K.
Using this column density of H$_2$, we find that the abundance of H$_2$O is 6.3 $\times$10$^{-9}$ for the envelope in a source size of 4$\arcsec$.
\citet{vdWiel13} also estimated the column density of H$_2$ for the envelope and outflow regions using {\it Herschel}/HIFI data of CO and they found that the column density of H$_2$ is $\sim$10$^{22}$ cm$^{-2}$ and 1.5 $\times$ 10$^{22}$ cm$^{-2}$ for the envelope and outflow, respectively.
Using these numbers, the abundance of H$_2$O is 1.7 $\times$10$^{-8}$ for the envelope and 2.8 $\times$ 10$^{-9}$ for the outflow.
Since our derived H$_2$O source size of 4$\arcsec$ is between the beam sizes of the infrared and submillimeter estimates for $N$(H$_2$), the most likely value 
for the H$_2$O abundance is between the above estimates. Since the outflow gas is more extended, the abundance estimate from the submillimeter CO data is probably the best.

In summary, our {\it Herschel}-HIFI observations indicate water abundances of $\sim$ 2$\times$10$^{-9}$$-$2$\times$10$^{-8}$ and a kinetic temperature of $\sim$ 60$-$230 K in the envelope.
These abundances are much lower than found from infrared (ISO) data (\S~1), which indicates that our observed emission comes mostly from the cold outer envelope.
Indeed, our derived water abundance is similar to the values of 5$\times$10$^{-10}$ to 4$\times$10$^{-8}$ found for the outer envelopes
of other high-mass protostars \citep{vdTak10, Chavarria10, Marseille10, Herpin12}, and also for low-mass protostellar envelopes \citep{Kristensen12}.
In contrast, the infrared absorption data are mostly sensitive to the warm inner envelope, and we expect this gas to emit primarily in the PACS lines rather than the HIFI lines. Indeed, using the {\it Herschel}/PACS instrument, \citet{Karska14} probe a warm water component which is similar to the gas seen with ISO.

For the outflow, we find a water abundance of $\sim$ 6$\times$10$^{-10}$$-$3$\times$10$^{-9}$ and a kinetic temperature of $\sim$ 70$-$90 K.
Despite the similar temperatures, the H$_2$O abundance in the outflow is a factor of 10 lower than in the AFGL 2591 envelope and also lower than found for other outflows, both from high-mass \citep{vdTak10, Emprechtinger13} and low-mass \citep{Bjerkeli12} protostars. The abundance estimate is uncertain through the adopted source size, but this effect is probably small as seen from a comparison of the column densities measured in absorption (Table 3) and emission (Table 4). The absorbing column is almost twice the emitting column, which suggests a source size of $\sim$20$\arcsec$ rather than $\sim$30$\arcsec$, assuming that the features arise in the same gas. The corresponding effect on the H$_2$O column density and abundance is only a factor of 2. However, the effect is probably smaller because the emission occurs over a smaller velocity range (from -15 to -5 km s$^{-1}$) than the absorption (from -24 to -5.5 km s$^{-1}$) which suggests that the features do not arise in the same gas. More likely, the water in the outflow lobes is affected by dissociating UV radiation, which also means that the H$_2$$^{18}$O lines are dominated by shocks at the interface between the jets and the envelope.

\section{Conclusions}
   \begin{enumerate}
      \item We present 14 rotational transitions of H$_2$O, H$_2$$^{17}$O, and H$_2$$^{18}$O
            toward the massive star-forming region AFGL 2591.
      \item We find redshifted water absorption from cold foreground clouds and blueshifted absorption from the outflow.
            Similar line features are found in W3 IRS5 \citep{Chavarria10} and DR21 \citep{vdTak10}
      \item We derived the o/p ratio using the ground-state lines of the main isotopologue and found that
            the o/p ratio is $\sim$2 in the cold foreground cloud, while  the o/p ratio is $\sim$3 in the warm protostellar envelope. Similar results are found toward Sgr B2(M) \citep{Lis10} and NGC 6334 I \citep{Emprechtinger10}.
            The inferred water abundances in the foreground clouds are consistent with ion-molecule chemistry.
     \item Radiative transfer models indicate that the envelope of AFGL 2591 is warm ($T_{\rm kin}$ $\sim$ 60$-$230 K). However, the low derived H$_2$O 
     abundance ($\sim$ 2$\times$10$^{-9}$$-$2$\times$10$^{-8}$) suggests that the H$_2$O line emission is dominated by the cold outer envelope where freeze-out of water onto dust grains is important. This apparent contradiction suggests that the water abundance in the protostellar envelope varies with radius.
      \item The water abundance in the outflow is $\sim$ 10$\times$ lower than in the envelope and in the outflows of other high- and low-mass protostars. Part of this difference may be due to beam size effect, but another possibility is the effect of dissociating UV radiation  \citep{vDishoeck13}. Compared to the envelope, the outflow lobes have a lower extinction, which leads to a higher UV radiation field and thus more rapid photodissociation. However, why this outflow has a lower H$_2$O abundance than those of other high-mass protostars is unclear. Models of UV-irradiated shocks are being developed to interpret the observed water abundances in protostellar outflows (M. Kaufman, priv. comm.).
   \end{enumerate}

The environment of AFGL 2591 has been the target of many observations from the ground and from space, but the {\it Herschel}/HIFI H$_2$O observations show the kinematics of this source (outflow, expanding envelope, foreground cloud) in more detail than any previous study.
This opportunity will be further explored with detailed radiative transfer models \citep{Hogerheijde00} in a future paper,
where we will estimate H$_2$O abundance profiles for a sample of high-mass protostellar envelopes.

\begin{acknowledgements}
The authors thank the referee for the careful and detailed report that helped to improve the paper. We also thank the editor, Malcolm Walmsley, for additional helpful comments.
We thank Kuo-Song Wang for the use of his population diagram code,
and Asunci{\'o}n Fuente and Timea Csengeri for useful comments on our manuscript.
HIFI has been designed and built by a consortium of institutes and university departments from across
Europe, Canada, and the US under the leadership of SRON Netherlands Institute for Space Research, Groningen,
The Netherlands, with major contributions from Germany, France and the US. Consortium members are:
Canada: CSA, U.Waterloo;
France: CESR, LAB, LERMA, IRAM;
Germany: KOSMA, MPIfR, MPS;
Ireland, NUI Maynooth;
Italy: ASI, IFSI-INAF, Arcetri-INAF;
Netherlands: SRON, TUD;
Poland: CAMK, CBK;
Spain: Observatorio Astron{\'o}mico Nacional (IGN), Centro de Astrobiolog{\'{\i}}a (CSIC-INTA);
Sweden: Chalmers University of Technology – MC2, RSS \& GARD, Onsala Space Observatory, Swedish National Space Board, Stockholm University – Stockholm Observatory;
Switzerland: ETH Z{\"u}rich, FHNW;
USA: Caltech, JPL, NHSC.
\end{acknowledgements}

\bibliographystyle{aa}                
\bibliography{AFGL2591_biblio}    

\Online
\begin{appendix}
\section{RADEX line ratio plot}

In Sects. 4.3 we calculate the line ratios of p-H$_2$O 2$_{11}-2_{02}$ and p-H$_2$O 2$_{02}-1_{11}$ using the non-LTE code RADEX \citep{vdTak07}. 
Figure A.1 shows the p-H$_2$O 2$_{11}-2_{02}$ and p-H$_2$O 2$_{02}-1_{11}$ line intensity ratio for kinetic temperatures between 20 K and 300 K and H$_2$ densities between 10$^3$ cm$^{-3}$ and 5$\times$10$^8$ cm$^{-3}$.

\begin{figure}
\includegraphics[width=8cm]{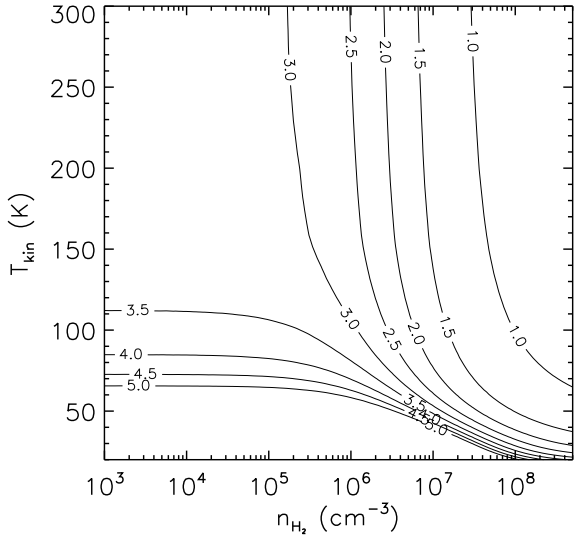}
      \caption{Line ratios of p-H$_2$O 2$_{11}-2_{02}$ and p-H$_2$O 2$_{02}-1_{11}$ as a function of kinetic temperature and H$_2$ density.}
       \label{fig7}
\end{figure}

\end{appendix}

\end{document}